\documentclass[pre,aps,twocolumn,superscriptaddress,longbibliography]{revtex4-1}

\usepackage[dvips]{graphicx}
\usepackage{amssymb,amsfonts,amsmath}
\usepackage{color}
\usepackage{ulem}

\newcommand{\rv}{{\mathbf r}}

\newcommand{\Jv}{{\bf J}}

\newcommand{\Fv}{{\bf F}}
\newcommand{\fv}{{\bf f}}
\newcommand{\ev}{{\bf e}}

\newcommand{\vel}{{\bf v}}

\newcommand{\tar}{{}}

\begin{document}

\title{Custom flow in overdamped Brownian Dynamics}

\author{Daniel de las Heras}
\email{Email: delasheras.daniel@gmail.com}
\affiliation{Theoretische Physik II, Physikalisches Institut, 
  Universit{\"a}t Bayreuth, D-95440 Bayreuth, Germany}
\author{Johannes Renner}
\affiliation{Theoretische Physik II, Physikalisches Institut, 
  Universit{\"a}t Bayreuth, D-95440 Bayreuth, Germany}
\author{Matthias Schmidt}
\email{Email: Matthias.Schmidt@uni-bayreuth.de}
\affiliation{Theoretische Physik II, Physikalisches Institut, 
  Universit{\"a}t Bayreuth, D-95440 Bayreuth, Germany}

\date{5 August 2018, revised version: 12 October 2018}

\begin{abstract}
When an external field drives a colloidal system out of equilibrium,
the ensuing colloidal response can be very complex and obtaining a
detailed physical understanding often requires case-by-case
considerations. In order to facilitate systematic analysis, here we
present a general iterative scheme for the determination of the unique
external force field that yields a prescribed inhomogeneous stationary
or time-dependent flow in an overdamped Brownian many-body
system. The computer simulation method is based on the exact one-body
force balance equation and allows to specifically tailor both gradient
and rotational velocity contributions, as well as to freely control
the one-body density distribution. Hence compressibility of the flow
field can be fully adjusted. The practical convergence to a unique
external force field demonstrates the existence of a functional map
from both velocity and density to external force field, as predicted
by the power functional variational framework. In equilibrium, the
method allows to find the conservative force field that generates a
prescribed target density profile, and hence implements the
Mermin-Evans classical density functional map from density
distribution to external potential.  The conceptual tools developed
here enable one to gain detailed physical insight into complex flow
behaviour, as we demonstrate in prototypical situations.
\end{abstract}

\maketitle

\section{Introduction}

The controlled application of an external field is a powerful means to
drive colloidal systems of mesoscopic suspended particles out of
equilibrium~\cite{LoewenRev,Rev2}. The resulting complex interplay of
the equilibrium properties of the system with the external
perturbation is already present in Perrin's pioneering work on
colloidal sedimentation~\cite{perrin}.  Gravitationally driven
colloidal suspensions~\cite{grav1,grav2,grav3} remain to this day
primary model systems for studying structure formation
phenomena. There is a large spectrum of different types of further
specific external influence on colloids, such as the response of
charged colloids to external electric fields~\cite{EF1,EF2}, and
magnetic field-induced transport of both diamagnetic and paramagnetic
colloidal particles~\cite{mag1,mag2} across a substrate, where the
colloidal motion was recently analysed in terms of the powerful
concept of topological protection against perturbation
\cite{topology}. The external magnetic fields in these setups varied
periodically in both space and time. Alternatively, exerting
shear-like flow on colloidal dispersions provides in-depth insights
into important effects in fundamental material science, such as shear
thickening~\cite{shear1,shear2} and glass
formation~\cite{shear3}. Furthermore, optical tweezers form powerful
and flexible tools for the generation of complex colloidal flow
patterns~\cite{ot1,ot2,ot3}.

In order to systematically study the response of a soft material to an
external perturbation, one typically first fixes the external field
and then studies the resulting colloidal motion under the effect of
the field. Certainly, this concept is compatible with our
understanding of a causal relationship between forces and the motion
that they generate.
If the external force field is static and conservative, then the
system will in general reach a new equilibrium state.  This response
of a complex system to such an external influence might be highly
non-trivial. As a result of the action of the external potential, the
system will in general become spatially inhomogeneous. In seminal
work, Mermin showed for quantum systems~\cite{PhysRev.137.A1441}, as
Evans subsequently did for the classical case~\cite{EvansDFT}, that a
functional inversion of the relationship between external potential
and one-body density distribution applies. Hence, reversing the above
``causal'' relationship, a unique mathematical map exists from the
one-body density distribution to the corresponding external
potential. This is an important and fundamental result of modern
Statistical Physics, which generalizes Hohenberg and Kohn's earlier
result for quantum ground states~\cite{PhysRev.136.B864}.
The functional relationship forms the basis of classical density
functional theory, which is used in practically all modern microscopic
theoretical treatments of spatially inhomogeneous systems~\cite{NDFT}.
Once the external potential is specified, the Hamiltonian is known
(the internal interactions remain unchanged) and hence {\it all}
equilibrium properties of the system are determined, and become
functional dependent on the density distribution.

In this work we address the functional map in non-equilibrium steady
and time-dependent states in overdamped Brownian many-body
systems. We first set the desired colloidal motion, as specified by
both the velocity field (or, equivalently, the one-body current
distribution) and the density profile, and then determine the specific
external force field that creates the prescribed motion in steady
state.  We develop and validate a numerical iterative method that
enables efficient and straightforward implementation of this task.

That the map from motion to external force field exists and that it is
unique follows formally from the power functional variational
principle in general time-dependent non-equilibrium \cite{PFT}. The
functional relationship has not, however, been explicitly demonstrated
in an actual many-body system. Here we provide the first such
demonstration, both for steady states and for time-dependent
  non-equilibrium. 

As a special case, we also apply the inversion method to equilibrium
systems. Here it allows to find the specific conservative external
force field that stabilizes a predefined density distribution. As flow
is absent and kinetic energy contributions are trivial in equilibrium,
this method also applies to inertial (i.e.\ Hamiltonian) systems.  An
alternative iterative numerical method that also implements the
functional map in equilibrium is presented in
Ref.~\cite{PRLsuperadiabatic}.

The conceptual progress in demonstrating the non-equilibrium inversion
map explicitly enables the practical solution to the problem of
generating tailor-made flow in complex systems. In the method that we
present, the sole requirement is, besides the ability to freely
control the external force field, to be able to measure the internal
one-body force density distribution. This is a readily available
quantity in many-body simulations, and it is conceivably also
accessible in experimental work.  We envisage that tailoring freely
flow on demand constitutes a powerful concept for the systematic study
of non-equilibrium physics. Here we investigate as a concrete
  model problem, the task of collecting particles in a certain region
  of space via a potential trap. The system is initially a homogeneous
  fluid, and we (i) speed up the natural dynamics by a factor of 2 and
  (ii) demonstrate that any unwanted effects due to superimposed
  external flow (e.g.\ due to convection or sedimentation in the
  system) can be fully cancelled.

The paper is organized as follows. In Sec.~\ref{SECtheory} we
describe how we obtain the iterative method, based on the exact
force balance relationship in overdamped Brownian dynamic, covering
both non-equilibrium steady states and equilibrium. In
Sec.~\ref{SECresults} we present results for several model
situations in which we custom-tailor the flow in a many-body system
of repulsive particles. Sec.~\ref{SECdiscussionConclusions} contains
a discussion and provides some conclusions. Details about particle
current sampling and convergence properties are given in the
appendix.

\section{Theory}
\label{SECtheory}
\subsection{Dynamical one-body force balance}
\label{SECforceBalance}
We consider a system of $N$ interacting Brownian particles in the
overdamped limit, where inertial effects are absent and we neglect
hydrodynamic interactions. The one-body density distribution
(``density profile'') at space point $\rv$ and time $t$ is given by
\begin{align}
  \rho(\rv,t) = 
  \Big\langle \sum_i \delta(\rv-\rv_i)  \Big\rangle,
  \label{EQdensityDefinition}
\end{align}
where the angles denote a statistical average, the expression inside
the angles is the microscopic density operator, with $\delta(\cdot)$
indicating the Dirac distribution, and $\rv_i$ denoting the position
of particle $i=1\ldots N$.  In the Fokker-Planck picture, the
information required for carrying out the average is encoded in the
many-body probability distribution $\Psi(\rv^N,t)$ of finding
microstate $\rv^N\equiv\rv_1\ldots\rv_N$ at time $t$. The average is
then defined as
\begin{align}
 \langle\cdot\rangle &= \int d\rv^N \cdot \Psi(\rv^N,t),
 \label{EQaverageFPdefinition}
\end{align}
where the integral runs over configuration space, i.e.\ each $\rv_i$
is integrated over the system volume.
The time evolution of $\Psi$ is determined by the Smoluchowski
equation $\partial \Psi/\partial t = -\sum_i \nabla_i\cdot\vel_i
\Psi$. Here the configuration space velocity $\vel_i$ of
particle~$i$ is given on the many-body level by the instantaneous
relation
\begin{align}
  \gamma \vel_i &= 
  - k_BT \nabla_i \ln \Psi,
  -\nabla_i u(\rv^N) + \fv_{\rm ext}(\rv_i,t),
  \label{EQvelocityManyBody}
\end{align}
where $\gamma$ is the friction constant against the implicit solvent,
$k_B$ is the Boltzmann constant, $T$ denotes absolute temperature,
$\nabla_i$ is the partial derivative with respect to~$\rv_i$, the
interparticle interaction potential is denoted by $u(\rv^N)$, and
$\fv_{\rm ext}(\rv,t)$ is the external force field, which in general
is position- and time-dependent. The three contributions on the right
hand side of \eqref{EQvelocityManyBody} correspond to thermal
diffusion (first term), deterministic motion due to interparticle
interactions (second term), and the externally imposed force (third
term).  This formulation of the dynamics is analogous to the Langevin
picture, where instead of \eqref{EQaverageFPdefinition}, the average
is taken over a set of stochastic particle trajectories for which a
random (position) noise provides the effects of thermal motion. The
corresponding discretized version is readily implementable in Brownian
Dynamics (BD) computer simulations (details of our implementation are
given in Sec~\ref{SECresults}).  Note that the configuration
  space velocity $\vel_i$ defined in \eqref{EQvelocityManyBody} is
  different from the average over the fluctuating velocity over
  realization of the noise, as represented in BD simulations.

We next supplement \eqref{EQdensityDefinition} by a corresponding
one-body current distribution, defined as
\begin{align}
  \Jv(\rv,t) &= \Big\langle
  \sum_i\delta(\rv-\rv_i)\vel_i
  \Big\rangle,
  \label{EQcurrentDefinition}
\end{align}
where $\vel_i$, at time $t$, is given
via~\eqref{EQvelocityManyBody}. We show in detail in appendix
\ref{ApCurrent} how a forward-backward symmetrical time derivative of
the particle positions can be used in BD to represent $\vel_i$ in
\eqref{EQcurrentDefinition}.

The density distribution \eqref{EQdensityDefinition} and the current
profile \eqref{EQcurrentDefinition} are linked via the continuity
equation,
\begin{align}
  \frac{\partial\rho(\rv,t)}{\partial t} &= 
  -\nabla\cdot\Jv(\rv,t).
  \label{EQcontinuity}
\end{align}
The many-body coupling in \eqref{EQvelocityManyBody} arises due to the
presence of the internal interaction potential $u(\rv^N)$. On the
one-body level, it is hence natural to define a corresponding internal
force density field via
\begin{align}
  \Fv_{\rm int}(\rv,t) = -\Big\langle
  \sum_i\delta(\rv-\rv_i)\nabla_i u(\rv^N)
  \Big\rangle.
  \label{EQinternalForceDensityAsAverage}
\end{align}
Here contributions to the average occur due to two effects, namely (i)
due to the bare value of the internal force field $-\nabla_i
u(\rv^N)$, but also (ii) due to the probability of finding particle
$i$ at the considered space point $\rv$, as measured by the delta
function.  Averages such as \eqref{EQinternalForceDensityAsAverage}
hence constitute microscopically resolved {\it force densities}.

By multiplying \eqref{EQvelocityManyBody} with $\delta(\rv-\rv_i)$,
summing over $i$, averaging according to
\eqref{EQaverageFPdefinition}, and identifying the one-body fields
\eqref{EQdensityDefinition}, \eqref{EQcurrentDefinition}, and
\eqref{EQinternalForceDensityAsAverage}, it is straightforward to show
that
\begin{align} 
  \gamma \Jv = -k_BT\nabla\rho + \Fv_{\rm int} +
  \rho\fv_{\rm ext},
  \label{EQforceDensityBalance}
\end{align}
which we use as a basis for the non-equilibrium inversion procedure.
Defining the microscopic velocity profile $\vel(\rv,t)$ simply as the
ratio between current profile and density profile,
\begin{align}
  \vel &= \Jv/\rho,
  \label{EQvelocityAsRatio}
\end{align}
allows us to rewrite \eqref{EQforceDensityBalance}, after division by
the density profile $\rho$, as
\begin{align}
  \gamma \vel &= -k_BT \nabla\ln\rho + \fv_{\rm int} +\fv_{\rm ext}.
  \label{EQforceBalanceNew}
\end{align}
Here the internal force field $\fv_{\rm int}$ is defined as the
internal force density \eqref{EQinternalForceDensityAsAverage}
normalized with the density profile, i.e.\ $\fv_{\rm
  int}(\rv,t)\equiv\Fv_{\rm int}(\rv,t)/\rho(\rv,t)$.

BD computer simulations allow straightforward access to the individual
contributions to the force balance relationship
\eqref{EQforceBalanceNew}. Sampling the density profile is
straightforward either using the traditional counting method or more
advanced techniques~\cite{Borgis2013,PRLcounting}.  The ideal
(diffusive) force field $-k_BT\nabla\ln\rho$ is readily obtained by
(numerical) differentiation of the density profile. The internal force
density field $\Fv_{\text{int}}$ can be sampled as an average over BD
realizations of the time evolution, or as an average over time when
investigating steady states; note that in BD one has direct access to
the internal force on the many-body level, $-\nabla_i u(\rv^N)$ in
\eqref{EQinternalForceDensityAsAverage}.

In typical applications, the external force field $\fv_{\rm
  ext}(\rv,t)$ is prescribed and $\rho(\rv,t)$ and $\vel(\rv,t)$
emerge as a result of the coupled many-body dynamics.  In the
following, we address the inverse problem of prescribing $\rho$ and
$\vel$ a priory and calculating the required form of $\fv_{\rm ext}$
that makes these fields stationary, such that the prescribed field
values are identical to the true dynamical averages of density
\eqref{EQdensityDefinition} and velocity \eqref{EQcurrentDefinition},
\eqref{EQvelocityAsRatio}.

\subsection{Inversion in nonequilibrium steady states}

\label{SECinversionNonequilibriumSteadyState}

Let $\rho_\tar(\rv)$ and $\vel_\tar(\rv)$ be the predefined stationary
(i.e.\ time-independent) ``target'' profiles for density and
velocity. In order to represent a valid steady state, the resulting
target current profile $\rho_\tar\vel_\tar$,
cf.~\eqref{EQvelocityAsRatio}, must be divergence-free,
$\nabla\cdot\rho_\tar\vel_\tar=0$, which follows from
\eqref{EQcontinuity} and represents a necessary condition on the
allowed set of target functions $\rho$, $\vel$. The external force
field $\fv_{\rm ext}(\rv)$ that makes the target profiles stationary
is obtained from first rearranging \eqref{EQforceBalanceNew} as
\begin{align} 
  \fv_{\text{ext}} = k_BT\nabla\ln\rho - \fv_{\text{int}} + \gamma\vel,
  \label{EQfextAsAnOutput}
\end{align}
where the internal force field,
$\fv_{\text{int}}(\rv)=\Fv_{\text{int}}(\rv)/\rho_\tar(\rv)$, is the
only unknown quantity on the right hand side, as $\rho$ and $\vel$ are
known input quantities. Here $\Fv_{\rm int}(\rv)$ is from
\eqref{EQinternalForceDensityAsAverage} in steady state.

In order to determine $\fv_{\rm int}$, and hence $\fv_{\rm ext}$ via
\eqref{EQfextAsAnOutput}, we proceed in two steps.  First we present a
fixed-point iterative scheme to solve \eqref{EQfextAsAnOutput}, in
which the $k$-th iteration step is defined via
\begin{align}
  \fv_{\text{ext}}^{(k)} = k_BT\nabla\ln\rho_\tar
  - \fv_{\text{int}}^{(k-1)} + \gamma\vel_\tar,
  \label{EQiterationStep}
\end{align}
where the targets, $\rho(\rv)$ and $\vel(\rv)$, are kept fixed for all
steps $k$.
Here $\fv_{\rm int}^{(k-1)}=\Fv_{\rm int}^{(k-1)}/\rho$, where
$\Fv_{\rm int}^{(k-1)}$ is the internal force density sampled in
the previous iteration step, $k-1$. Data for $\Fv_{\rm int}^{(k-1)}$
was obtained from BD sampling under the prescribed external force
field $\fv_{\text{ext}}^{(k-1)}$. In order to initialize the iteration
scheme, we set the external force field at step $k=0$ simply as
\begin{align}
  \fv_{\text{ext}}^{(0)} &=  k_BT\nabla\ln\rho_\tar + \gamma\vel_\tar, 
  \label{EQiterationStart}
\end{align}
which is the exact external force field for the case of an ideal
gas. Prescribing \eqref{EQiterationStart} allows to sample $\Fv_{\rm
  int}^{(0)}$ in BD, and then use $\fv_{\rm int}^{(0)}$ as the
required input for iteration step $k=1$ in~\eqref{EQiterationStep}.
This completes the description the functional inversion.

At each iteration step we also sample both the one-body density and
one-body current profiles, $\rho^{(k)}$ and $\Jv^{(k)}$; details on
how to sample the current in BD are provided in
Appendix~\ref{ApCurrent}.  As a criterion for judging the eventual
convergence of $f_{\rm ext}^{(k)}$ to the real external force field
that makes the target density and current profiles stationary,
i.e.\ the solution of \eqref{EQfextAsAnOutput}, we use the difference
between the target and the sampled profiles at step $k$, i.e.,
\begin{eqnarray}
  \Delta\rho=\int d\rv\big(\rho_\tar(\rv)
  -\rho^{(k)}(\rv)\big)^2/V<\epsilon_1,\label{errorDensity}\\
  \Delta J= \int d\rv\big\lvert\Jv_\tar(\rv)
  -\Jv^{(k)}(\rv)\big\rvert^2/V<\epsilon_2,\label{errorCurrent}
\end{eqnarray}
where $V=\int d\rv$ is the system volume, and
$\epsilon_1,\epsilon_2>0$ are small tolerance parameters. Numerical
details of our implementation are given 
in Appendix~\ref{ApB}.

We find that in practice the iteration method converges reliably in
all cases considered; results are shown below in
Sec.~\ref{SECresults}. That a solution exists for $\fv_{\rm ext}$ and
that it is unique are nontrivial properties of our scheme. We expect
existence and uniqueness to hold, however, based on the power
functional variational framework~\cite{PFT}, which is a novel approach
for the statistical description of the dynamics of many-body
systems. The central object of power functional theory (PFT) is a
``free power'' functional $R_t[\rho,\Jv]$ of the one-body density and
current or analogously, viz.\ \eqref{EQvelocityAsRatio}, of the
density and the velocity field. $R_t$ has units of energy per time
(power) and plays a role analogous to the free energy functional (as
detailed below) in equilibrium. It consists of an ideal gas
contribution ($W_t^{\rm id}$), an excess (over ideal gas) part due to
the internal interactions ($W_t^{\rm exc}$) and an external power
($X_t$) contribution, according to $R_t=W_t^{\rm id}+W_t^{\rm
  exc}-X_t$.

PFT implies that $\fv_{\rm int}$ is a unique functional of density and
current distributions, or equivalently of density and velocity
profile. In particular, $\fv_{\rm int}$ can be expressed as a
functional derivative of the intrinsic excess (over ideal gas) free
power functional $W_t^{\rm exc}[\rho,\Jv]$, as
\begin{align}
  \fv_{\rm int}([\rho,\vel],\rv,t) &=
  -\frac{\delta W_t^{\rm exc}[\rho,\Jv]}{\delta \Jv(\rv,t)},
  \label{EQfintAsDerivative}
\end{align}
where the density distribution $\rho$ is kept fixed upon the
variation, at fixed time $t$.

Typically, one would split further into adiabatic and superadiabatic
contributions, $W_t^{\rm exc} = \dot F_{\rm exc}[\rho]+P_t^{\rm
  exc}[\rho,\Jv]$, where $\dot F_{\rm exc}[\rho]$ is the time
derivative of the equilibrium excess (over ideal gas) Helmholtz free
energy functional, and $P_t^{\rm exc}[\rho,\Jv]$ is the superadiabatic
contribution, which describes the genuine nonequilibrium effects. This
splitting offers great advantages in terms of the classification of
the different types of forces that occur, but it is not required for
our present purposes. We rather work directly with $\fv_{\rm int}$.
Recall that this is directly accessible via
Eq.~\eqref{EQinternalForceDensityAsAverage} in BD simulations.

The fact that $\fv_{\rm int}$ is generated from a current-density
functional, via functional differentiation \eqref{EQfintAsDerivative},
implies that the force field itself is a functional of density and
current (or velocity profile).  Hence \eqref{EQfintAsDerivative}
constitutes a unique map from density and velocity to the internal
force field,
\begin{align}
  \fv_{\rm int}(\rv) &= \fv_{\rm int}([\rho,\vel],\rv,t),
\end{align}
where the right hand side is the force field functional
\eqref{EQfintAsDerivative} evaluated at the target profiles $\rho$ and
$\vel$; the left hand side is the corresponding (hitherto unknown)
specific form of the internal force on the left hand side of
\eqref{EQfextAsAnOutput}.  
Hence by inserting \eqref{EQfintAsDerivative} into
\eqref{EQfextAsAnOutput} we obtain the explicit form
\begin{align}
  \fv_{\rm ext} &= k_BT\nabla\ln\rho - \fv_{\rm int}[\rho,\vel]
  +\gamma \vel,
  \label{EQfextExplicitFromFunctional}
\end{align}
with the iteration procedure \eqref{EQiterationStep} and
\eqref{EQiterationStart} being a practical scheme for evaluating the
right hand side.  Note that \eqref{EQfextExplicitFromFunctional} is an
explicit expression for $\fv_{\rm ext}$; no hidden dependence on
$\fv_{\rm ext}$ occurs on the right hand side. Recall that from
\eqref{EQfintAsDerivative}, the internal force field depends solely on
the ``kinematic'' fields $\rho$ and $\vel$, but not on the external
force field. This completes the proof. Before presenting
results, we revisit the equilibrium case.

\subsection{Inversion in equilibrium}
In equilibrium, the velocity profile is identically zero, and we
therefore can simply set the target $\vel_\tar(\rv)\equiv 0$ and
prescribe $\rho(\rv)$ in order to find the corresponding external
force field $\fv_{\rm ext}(\rv)$. The external force field is
necessary of conservative, gradient form, $\fv_{\rm ext}(\rv)=-\nabla
v_{\rm ext}(\rv)$, where $v_{\rm ext}$ is the external potential
energy.  Clearly, there is no dependence on time, and, as we show, one
only needs to carry out equilibrium averages. Hence the method also
applies to Hamiltonian systems, as the kinetic contributions are
trivial.

In equilibrium we can simplify \eqref{EQfextAsAnOutput} to obtain
\begin{align} 
  \fv_{\rm ext} &= k_BT\nabla\ln\rho_\tar - \fv_{\text{int}}.
  \label{EQforceBalanceEquilibrium}
\end{align}
which constitutes an explicit expression for the specific external
force field that generates the given target profile $\rho_\tar$ in
equilibrium.

As it is the case in nonequilibrium steady state, the internal force
is unknown, but it can be found iteratively. The iteration step is
\begin{align} 
  \fv_{\rm ext}^{(k)}(\rv)
  &= k_BT\nabla\ln\rho_\tar(\rv) - \fv_{\text{int}}^{(k-1)}(\rv).
  \label{iteeq}
\end{align}
and the external force is initialized with the exact solution of an
ideal gas,
\begin{align} 
  \fv_{\rm ext}^{(0)}(\rv)
  &= k_BT\nabla\ln\rho_\tar(\rv), 
  \label{EQiterationStartEquilibrium}
\end{align}
We then sample $\fv_{\text{int}}^{(0)}$ in equilibrium, under the
external force
$\fv_{\rm ext}^{(0)}$, and then iterate, on the basis of
\eqref{iteeq}, until the difference between the target and the sampled
density profiles is small, cf.~\eqref{errorDensity}. As only the
internal force and the density profiles are required, the sampling can
be performed using BD or molecular dynamics simulations. If one wishes
to use the Monte Carlo method, then one needs the actual value of the
potential $v_{\text{ext}}^{(k)}$ instead of the force, $\fv_{\rm
  ext}^{(k)}=-\nabla v_{\text{ext}}^{(k)}$. In systems that
effectively depend on only one coordinate, say $x$, the potential at
each iteration can be easily obtained from the internal force profile,
by performing a one-dimensional spatial integral
\begin{align} 
  v_{\text{ext}}^{(k)}(x)=-k_BT\ln\rho_\tar(x) 
  + \int dx f_{\text{int}}^{(k-1)}(x).\label{ulala}
\end{align}
In two- and three-dimensional systems a line integral or, more
generally, the use of an inverse operator $\nabla^{-1}$ is required to
obtain the potential from the force field. Hence the situation is
similar to the one addressed in modern ``force sampling'' methods that
yield the density profile \cite{Borgis2013,PRLcounting}.

That the method converges to a unique external potential is guaranteed
by the Mermin-Evans functional map from density profile to external
potential \cite{PhysRev.137.A1441,EvansDFT}. In particular, the
internal force field is obtained as a functional derivative of the
excess free energy functional via
\begin{align}
  \fv_{\rm int}([\rho],\rv) &= 
  -\nabla \frac{\delta F_{\rm exc}[\rho]}{\delta \rho(\rv)}.
\end{align}
Inserting this into \eqref{EQforceBalanceEquilibrium} yields
\begin{align}
  \fv_{\rm ext} &= k_BT\ln\rho 
  +\nabla\frac{\delta F_{\rm exc}[\rho]}{\delta\rho},
  \label{EQfextExplicitFromFunctionalEquilibrium}
\end{align}
which is an explicit expression for the external force field, given
$\rho$ as an input, in analogy to the nonequilibrium case,
\eqref{EQfintAsDerivative} and
\eqref{EQfextExplicitFromFunctional}.
For completeness, and briefly,
\eqref{EQfextExplicitFromFunctionalEquilibrium} is formally obtained
from the more general \eqref{EQfintAsDerivative} and
\eqref{EQfextExplicitFromFunctional} by observing that in equilibrium
$\delta W_t^{\rm exc}/\delta \Jv=\delta \dot F_{\rm exc}/\delta \Jv$,
where $\dot F_{\rm exc}=\int d\rv \Jv\cdot\nabla \delta F_{\rm
  exc}/\delta \rho$, cf.~\cite{PFT}.

We next clarify the relationship to the method of
Ref.~\cite{PRLsuperadiabatic}. Note that at any step in the iteration,
given the external force field $\fv^{(k-1)}_{\rm ext}$, the internal
force field may be written as
\begin{align}
  \fv^{(k-1)}_{\rm int}= k_BT \nabla 
  \ln(\rho^{(k-1)}) -\fv^{(k-1)}_{\rm ext},
  \label{EQfromReferee1}
\end{align}
in terms of the density distribution $\rho^{(k-1)(\rv)}$ at
equilibrium with that external force. Then Eq.~\eqref{iteeq} may be
written as the change in the force, along the iterative procedure,
\begin{align}
  \fv^{(k)}_{\rm ext} - \fv^{(k-1)}_{\rm ext} 
  = k_BT \nabla \ln( \rho/\rho^{(k-1)}),
  \label{EQfromReferee2}
\end{align}
that vanishes when the target density is achieved,
$\rho=\rho^{(k-1)}$.  The integration of \eqref{EQfromReferee2}, to
get the change in external potential, and the linear expansion
\begin{align}
  \ln(\rho/\rho^{(k-1)}) = -(\rho^{(k-1)}- \rho)/\rho +\ldots
  \label{EQfromReferee3}
\end{align}
give precisely the method used in Ref.~\cite{PRLsuperadiabatic}.

\subsection{Inversion in time-dependent nonequilibrium}
\label{SECinversionTimeDependent}

In order to perform the inversion in time-dependent nonequilibrium, we
carry out the procedure of
Sec.~\ref{SECinversionNonequilibriumSteadyState} at a discretized
sequence of (coarse-graining) times $t_{\text {cg}}$ during the
time evolution. The method propagates the system forward in time, in
sync with the target time evolution. At each coarse-graining time step the required
external force field is obtained (via iteration) such that the
prescribed target density $\rho(\rv,t_{\text{cg}})$ and velocity field
$\vel(\rv,t_{\text{cg}})$ are identical to their respective values in the target
time evolution of the system. We interpolate linearly the values for
the external force field between two consecutive times, which we find sufficient
for the test cases presented below.

In detail, at each coarse-graining time step $t_{\text{cg}}$ we iterate the value of the
external field according to
\begin{align}
  \fv_{\rm ext}^{(k)}(\rv,t_{\text{cg}}) &=
  k_BT\nabla\rho(\rv,t_{\text{cg}})
  -\fv_{\rm int}^{(k-1)}(\rv,t_{\text{cg}})
  +\gamma \vel(\rv,t_{\text{cg}}),
  \label{EQtimeDependentIteration}
\end{align}
where $\rho(\rv,t)$ and $\vel(\rv,t)$ are the target fields, which
enter via their values at time $t_{\text{cg}}$. The time $t_{\text{cg}}$ is kept fixed
under the iteration $k\to k+1$ described by
\eqref{EQtimeDependentIteration}. For the first time step we initialize the external
force using the exact ideal gas solution, Eq.~\eqref{EQiterationStart}. For the 
subsequent time steps we initialize the iterative scheme using the solution of the
previous time step:
\begin{equation}
\fv_{\text{ext}}^{(0)}(\rv,t_{\text{cg}})=\fv_{\text{ext}}(\rv,t'_{\text{cg}}),
\end{equation}
where $t'_{\text{cg}}$ indicates the time step previous to $t_{\text{cg}}$.

The iteration method proceeds forward in time. In order to correctly
account for memory effects, the many-body dynamics evolves according
to continuous, valid trajectories over the entire time-dependent
dynamics. In the BD simulations, this requires to start a new coarse-graining time step
using the many-body configuration(s) obtained at the end of
previous coarse-graining time step. At the end of the process, the entire field $\fv_{\rm
  ext}(\rv,t)$ is known, and as consistency check, can be input into a
``bare'' non-steady BD run, in order to validate that the targets
$\rho(\rv,t)$ and $\vel(\rv,t)$ are met during the entire course of
time.

\section{Results}
\label{SECresults}
In the following we demonstrate that the straightforward
application of the method allows to cast new light on fundamental
physical effects by studying a two-dimensional model fluid
of Brownian particles interacting via the common
Weeks-Chandler-Anderson potential \cite{truncatedLJ}, i.e.\ a purely
repulsive, truncated-and-shifted Lennard-Jones (LJ) pair
potential~\cite{truncatedLJ},
\begin{equation}
\phi(r)=\left\{
        \begin{array}{ll}
                4\epsilon\left[\left(\frac\sigma r\right)^{12}-\left(\frac\sigma r\right)^{6}+\frac14\right] & \text{if } r < r_c\\
                0 & \text{otherwise,}\\ 
        \end{array} 
\right.
\end{equation}
where the parameters $\epsilon$ and $\sigma$ set the energy and length
scales, respectively, and $r$ indicates the center-center distance of
the particle pair. The cutoff distance $r_c=2^{1/6}\sigma$ is located
at the minimum of the standard LJ potential, and hence the interaction
is purely repulsive.

The particles are in a square box of length $L$ with periodic boundary
conditions along both directions. Using the standard Euler algorithm,
the Langevin equation of motion is integrated in time via
\begin{align}
\rv_i(t+\Delta t)=\rv_i(t)+\frac{\Delta t}{\gamma}
   [-\nabla_i u(\rv^N)
  + {\bf f}_{\rm ext}(\rv_i,t)]
 + \boldsymbol\eta_i(t),
\label{EQBD} 
\end{align}
where $\boldsymbol\eta_i$ is a delta-correlated Gaussian random
displacement with standard deviation $\sqrt{2\Delta t k_BT/\gamma}$ in
accordance with the fluctuation-dissipation theorem. Here, $\Delta t$
is the integration time step that we set to $\Delta t/\tau = 10^{-4}$
with $\tau=\sigma^2\gamma/\epsilon$ the unit of time; the friction
constant is set to $\gamma=1$.

\subsection{Effective one-dimensional system}
A considerably large class of nonequilibrium situations is effectively
of one-dimensional nature, where both the density profile and the
current distribution depend only on a single coordinate, say $x$, and
the flow direction is along the $x$-axis (i.e.\ with no shear motion
occurring). Then the steady state condition reduces to the requirement
of the current being constant, $\Jv(x)=J_0\ev_x$, where $J_0=\rm
const$ and $\ev_x$ is the unit vector in the $x$-direction. Hence from
\eqref{EQvelocityAsRatio} the velocity and the density profile possess
a reciprocal relationship: $v(x)=J_0/\rho(x)$.

\begin{figure}
\includegraphics[width=0.90\columnwidth]{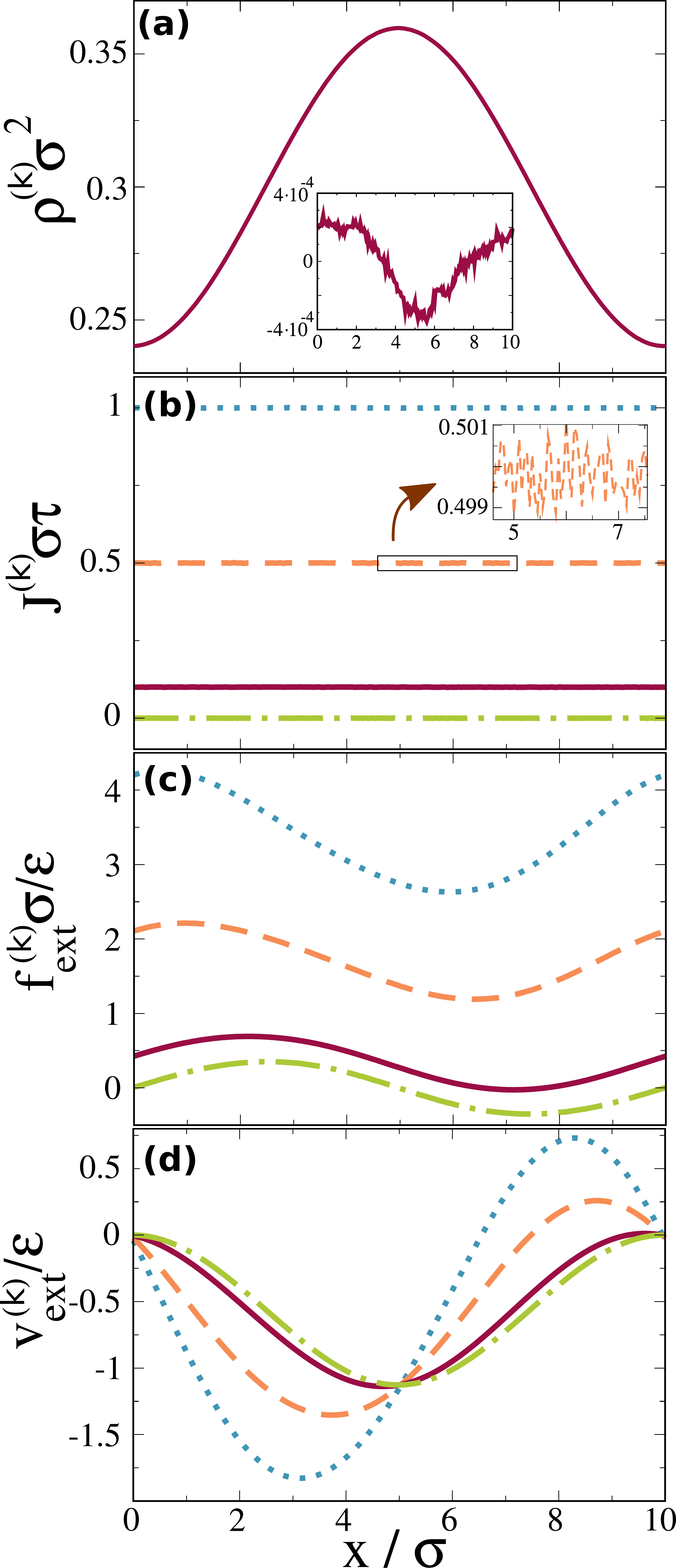}
\caption{One-body density (a) and current profiles (b), external force field (b), and
  external potential (c) as a function of the $x$-coordinate in
  a system with target density profile 
  $\rho_\tar(x)=c_1\sin^2(\pi x/L)+c_2$ with $c_1\sigma^{2}=0.12$ and
  $c_2\sigma^2=0.24$ after $k=40$ iterations.
  The inset in (a) shows the difference between target and sampled density profiles.
  Results are shown for different values of the target current: $J_0\sigma\tau=1$
  (blue dotted line), $J_0\sigma\tau=0.5$ (orange dashed line),
  $J_0\sigma\tau=0.1$ (violet solid line), and $J_0\sigma\tau=0$
  (green dot-dashed line), which is in equilibrium. The inset in (b)
  is a close-view for $J_0\sigma\tau=0.5$. 
  Two-dimensional system with $N=30$ particles in a periodic box of
  side $L/\sigma=10$ at $k_BT/\epsilon=1$. Data 
  obtained by averaging over $25$ BD realizations (MC
  realizations in equilibrium).}
\label{fig1}
\end{figure}

We first study such an effective one-dimensional problem with $N=30$
particles in a square simulation box of side length $L/\sigma=10$.  We
choose the target density profile to contain a single nontrivial
Fourier component, that modulates the homogeneous fluid,
\begin{align}
  \rho_\tar(x)=c_1\sin^2(\pi x/L)+c_2,\label{targetrho1}
\end{align}
with $c_1\sigma^{2}=0.12$ and $c_2\sigma^2=0.24$ such that $\int
dr\rho(\rv)=N$. See the density profile in Fig.~\ref{fig1}(a). The
temperature is set to $k_BT/\epsilon=1$.

Our inversion method facilitates the study of fundamental aspects of
driven systems. As an illustrative example, we construct a family of
steady states that share the same density profile,
cf. Eq.~\eqref{targetrho1}, but possess different values $J_0$ of the
(constant) target current. In Appendix~\ref{ApB} we describe numerical
details of the concrete implementation of the iterative procedure.

We show in Fig.~\ref{fig1} the external force field required to make
the target density profile~\eqref{targetrho1} stationary, for a range
of different values of $J_0$. In Figs~\ref{fig1}(a) and (b) the final
converged density and current profiles are shown; these are
  indeed (numerically) identical to their targets. We consider four
steady states with values of the current $J_0\sigma\tau=0$
(equilibrium), $0.1,0.5$ and $1$. The specific external force field
required to produce each such steady states is depicted in
Fig.~\ref{fig1}(c) for all four cases. The force fields can be
represented as the sum of a spatially constant force offset plus a
conservative potential contribution. The constant offset drives the
particle flow and it can be calculated as the spatial average of
  the total external force field. The conservative term generates the
density modulation. As expected, in the equilibrium case
  ($J_0=0$) only the conservative term is present, and we find that
  the spatial average of the total external force vanishes. In
Fig.~\ref{fig1}(d) we show the external potential $v_{\text{ext}}(x)$
that generates the conservative force contribution.  As a convention,
we have introduced an (irrelevant) shift of the energy scale, such
that $v_{\rm ext}=0$ at $x=0$ for all four cases. As expected, in
equilibrium $v_{\text{ext}}$ possesses a minimum at the location of
the density peak. It turns out that in order to keep the density
profile unchanged upon imposing the constant flux of particles in the
$x$-direction, the external potential changes its shape very
substantially. Both the minimum and the maximum move towards
smaller values of $x$, i.e.\ against the direction of the flow, upon
increasing $J_0$ (note the periodicity in $x$). Clearly, this
behaviour is a direct consequence of keeping the density profile
constant while increasing the flow through this density
``landscape''. In order to rationalize this effect, consider first
the case where an external potential generates the the density
profile, Fig.~\ref{fig1}(a), in equilibrium. If we now switch on a
an additional constant (positive) external force contribution, the
result will be a particle flow and the density profile will respond
by shifting the density peak in the direction of the flow (results
not shown). In our system the density profile is rather kept
constant and the shifting of the density peak needs to be cancelled
by the external conservative field, which hence necessarily develops
the observed shift in the direction opposite to the flow. 
Besides quantifying the positional shift, cf.\ Fig.~\ref{fig1}(d),
we also observe a marked increase in the amplitude of the external
potential contribution; hence stronger ``ordering'' forces,
$-\nabla v_{\rm ext}$, are required in order to overcome the
homogenizing effect of the flow.

\subsection{Two-dimensional system}
The iteration scheme is general and it
is not restricted to effectively one-dimensional inhomogeneous
systems. As a proof of concept, we construct the external force field
that makes a two-dimensional density profile stationary. We choose the
target velocity field to be
\begin{equation}
\vel_\tar(x,y) = \left(
        \begin{array}{c}
                d_1 \sin(2\pi y / L) \\
                d_2 \\ 
        \end{array}
        \right).
\label{veltarget}
\end{equation}
with $d_1,d_2=\rm const$.  
As above, a companion target density profile cannot be
chosen arbitrarily, since the resulting current must satisfy the
steady state condition, $\nabla\cdot\rho\vel=0$.  Given that
\eqref{veltarget} is divergence-free, $\nabla\cdot\vel_\tar=0$, the
steady state condition reduces to $\vel_\tar\cdot\nabla\rho_\tar=0$.
As an immediate first choice, we set
\begin{align}
  \rho(x,y)=N/L^2=\rm const,
  \label{EQconstantTargetDensity}
\end{align}
which trivially satisfies the steady state condition. Note that
\eqref{veltarget} and \eqref{EQconstantTargetDensity} represent a
conceptually highly interesting case of a homogeneous, bulk-fluid-like
one-body density distribution, with ``superimposed'' flow
\eqref{veltarget} that is fully inhomogeneous on microscopic length
scales.

Furthermore, as a second choice together with \eqref{veltarget}, we
consider the target density profile
\begin{align}
  \rho_\tar(x,y) &= N/L^2+a_0\cos\left(2\pi x/L 
  + Y\right)\label{targetrho},\\
  Y &=d_0\cos(2\pi y/L)\label{targetrho2},
\end{align}
such that $Y(y)$ is a spatially modulating function of the given form,
$a_0$ is a constant such that $a_0<N/L^2$ (in order to ensure that the
$\rho>0$, and $d_0=d_1/d_2$.
Since $\nabla\rho_\tar$ is perpendicular to $\vel$ for
all $\rv$, it is straightforward to show that \eqref{veltarget},
\eqref{targetrho} and \eqref{targetrho2} also define a valid steady
state.

\begin{figure*}
\includegraphics[width=0.90\textwidth]{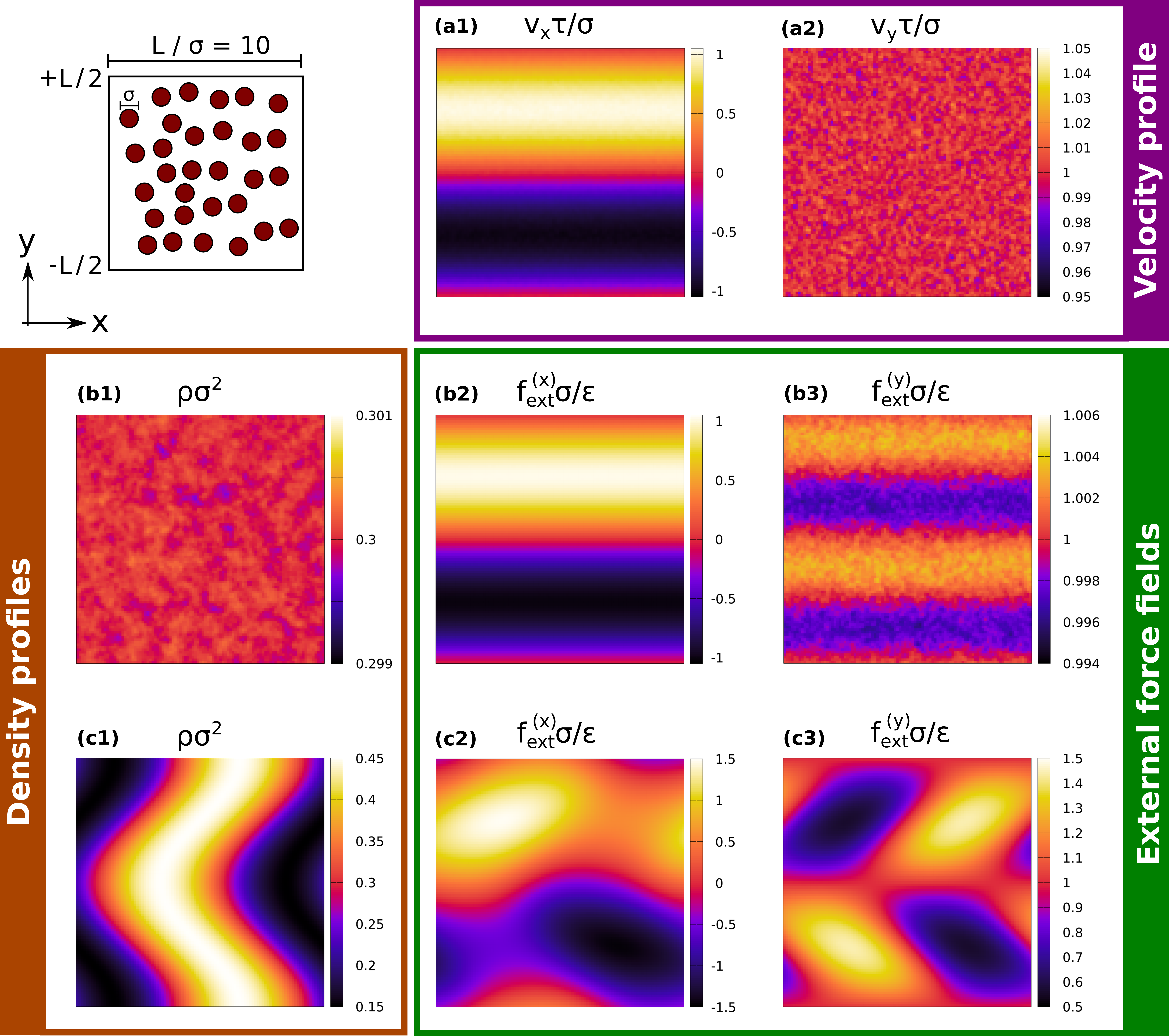}
\caption{$x-$component (a1) and $y-$component (a2) of the velocity profile sampled after $k=40$ iterations using BD simulations.
(b1) Sampled density profile for the steady state with constant density profile. $x-$ (b2) and  $y-$ (b3) components of the
external force field that produces the steady state with constant density. (c1) Sampled density profile for the steady state
with inhomogeneous density profile. $x-$ (c2) and  $y-$ (c3) components of the external force field that generates
the steady state with inhomogeneous density. In both steady states we set $N=30$, $L/\sigma=10$, and $k_BT/\epsilon=0.5$.
The bin size is set to $0.05\sigma$ in both directions and the origin or coordinates 
is located in the middle of the box. Results are averages over $100$ BD realizations.}
\label{fig2}
\end{figure*}

For both target states (constant and non-constant density profiles) we
use the inversion method to find the external force fields that
renders a stationary situation. We set $N=30$, $L/\sigma=10$, and
$k_BT/\epsilon=0.5$.  For the target velocity profile, we set 
  $d_1=d_2=\tau/\sigma$ in Eq.~\eqref{veltarget}. For the
inhomogeneous density profile we set $a_0=0.5N/L^2$ in
Eq.~\eqref{targetrho}. 

The two Cartesian components of the velocity profile, obtained after
$40$ BD iterations of the inversion method, are shown in
Fig.~\ref{fig2} (a1) and (a2). The sampled velocity and density
profiles coincide with the target profiles within the imposed
numerical accuracy. The sampled density profiles are shown in
Fig.~\ref{fig2}(b1) (constant density profile) and Fig.~\ref{fig2}(c1)
(inhomogeneous density profile). The corresponding external force
fields are presented in panels (b2-b3) and (c2-c3) of Fig.~\ref{fig2}.

For the case of constant density, the $x$ component of the external force field $f_{\text{ext}}^{(x)}$, see Fig.~\ref{fig2}(b2),
is very similar in shape and magnitude to the $x$ component of the velocity profile, Fig.~\ref{fig2}(a1).
Given  that the friction coefficient is set to $\gamma = 1$, this means that $f_{\text{ext}}^{(x)}$
generates the flow in the $x$ direction (there are small differences between $f_{\text{ext}}^{(x)}$ and $\gamma v_x$ related
to the $x$ component of the internal force field). The $y$ component of the external force field, shown in Fig.~\ref{fig2}(b3), 
shows a small deviation from an average value which is consistent with the value of the flow in~$y$.
This deviation is expected, since we have imposed a constant density
profile, and hence the external force has to balance the migration
force~\cite{shearmigration,shearmigration1} that results from the
shear field imposed by $v_x$. The $y$ component of the external force
is inhomogeneous but the density profile is constant. Hence, the
internal force must cancel the action of $f_{\text{ext}}^{(y)}$. This
is a purely superadiabatic effect~\cite{PRLstructural}, which is completely
neglected in the widely-used dynamical density functional
theory (DDFT) ~\cite{DDFT} which rather predicts internal forces to vanish for
situations of constant density. Extended versions of DDFT have been recently proposed to 
try to overcome these limitations, see e.g.~\cite{DDFTBrader,DDFTBrader2}.

The target velocity profile is effectively one-dimensional, and the
target density profile is constant. As a result the external force is
also effectively one-dimensional. This is not the case when the target
density profile is inhomogeneous. Then the $x$ and $y$ components of
the external force field depend on both coordinates, see
Fig.~\ref{fig2}, panels (c2) and (c3). Now the external force field
generates the flow and also sustains the density gradient. Clearly,
the required force field, which generates the fully inhomogeneous
steady state, is very complex, and simple physical reasoning, such as
we could rely on in the former two cases, is insufficient to obtain
even a qualitative, let alone (semi-)quantitative rationalization of
the occurring physics.

\subsection{Dynamic confinement}

While the above examples demonstrate custom flow for steady states, we
next turn to its implementation for full (time-dependent)
nonequilibrium situations, as laid out in
Sec.~\ref{SECinversionTimeDependent}. We hence aim to show that the
concept is general and valid even for complex dynamics.

\begin{figure*}
\includegraphics[width=0.95\textwidth]{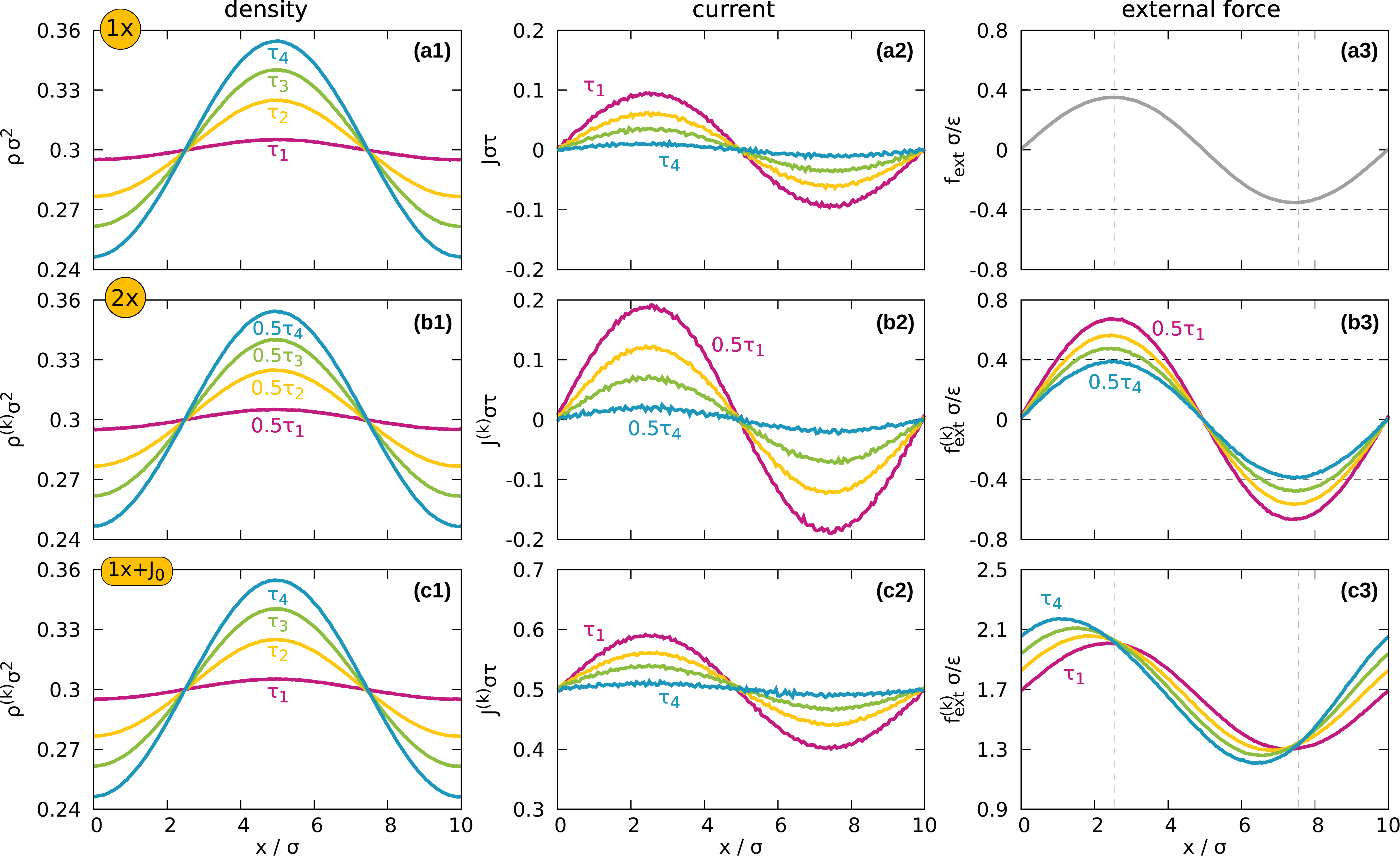} \caption{Inversion in
  full nonequilibrium as demonstrated by dynamics of
  confinement. Density profile (left column), current profile (middle
  column), and external force field (right column) as a function of
  $x$ for three different situations. (a) Time evolution of the
  system, density (a1) and current (a2), after switching on a static
  external field (a3). At $t=0$ the system is in equilibrium with
  vanishing external field. Results at four times are shown:
  $\tau_1/\tau=0.08$, $\tau_2/\tau=0.48$, $\tau_3/\tau=1.0$, and
  $\tau_4/\tau=2.2$. At $\tau_4$ the system is near equilibrium with
  the applied external field (a3). Profiles are obtained by averaging
  over $\sim10^8$ different trajectories. In panels (b) we show a
  system that evolves following the same dynamics as in (a) but two
  times faster.  The current (b2) is therefore twice as large as the
  current in the original system, and the required external field (b3)
  is time dependent. A movie showing a system that evolves three times
  faster is presented in the Supplemental Material \cite{SuppMat}.
  Panels (c) show the time evolution in a system that reproduces the
  same dynamics as (a) but with a global motion towards the right
  (note that the spatial average of the current (c2) at any time is
  $J_0\sigma\tau=0.5$).  The required external field (c3) is
  time-dependent. The horizontal and vertical dashed lines in the
  plots of the external field are drawn to help the comparison between
  systems. In all cases we set $N=30$, $L/\sigma=10$, and
  $k_BT/\epsilon=1.0$.}
\label{fig3}
\end{figure*}

As a prototypical situation, we address the time evolution of a
system, which in its initial state is a homogeneous equilibrium fluid,
(with no external field acting in this initial state). At time $t=0$ we
switch on a conservative external field, which represents the potential
trap shown Fig.~\ref{fig1} panels (c) and (d) for the equilibrium case
(green dot-dashed-line). Hence, the external
force induces migration of particles towards the center of the system,
as shown in panels (a) of Fig.~\ref{fig3} for the density (a1) and the
current (a2).

The external field is static for $t>0$, see panel (a3), and its
influence evolves the system from the homogeneous state to a confined
state that features a well-defined, peaked density modulation,
Fig.~\ref{fig3}(a1). After only few Brownian times, a new equilibrium
state is reached. The particle current almost vanishes already at time
$\tau_4/\tau=2.2$, see Fig.~\ref{fig3}(a2).

Using the time-dependent version of the custom flow method described
in Sec.~\ref{SECinversionTimeDependent}, we chose to determine the
time- and position-dependent external force field,
$f_{\text{ext}}(t,x)$, that speeds up the dynamics of the system by a
factor $\alpha>1$.  That is, we find a system that evolves through the
same temporal sequence of density profiles as those in
Fig.~\ref{fig3}(a), but doing so at a rate which $\alpha$ times
faster. Hence in the new ``fast forward'' system the density profile
$\rho_\alpha$ at time $t$ is the same as the density profile in the
original system at time $\alpha t$. The current in the new system must
be $\alpha$ times the current in the old system due to the continuity
equation. That is, in the new system:
\begin{align}
  \rho_\alpha(t,x) &=\rho(\alpha t,x),\nonumber\\
  J_\alpha(t,x) &= \alpha J(\alpha t,x).
\end{align}
In order to find the external field that induces the desired target
dynamics, we discretize the time evolution at intervals $\Delta
t_{\text{cg}}/\tau=0.01$, i.e.\ on a scale that is $10^2$ times larger
than the time step of the BD simulation $\Delta t/\tau=10^{-4}$. At
each coarse-graining time $t_{\rm cg}$ we run an iterative process to
find the desired external field at that time. We use a linear
regression to approximate the external field at every time $t$ between
two consecutive coarse-graining times. The imposed coarse-graining time
is a good compromise between accuracy and computational time. Since
the external field does not vary profusely during one time interval
only a few iterations ($<10$) are required at each $t_{\text{cg}}$. At
each iteration we average over $10^6$ trajectories. Finally, we
average the results over $50$ independent simulation runs.

The dynamics of the system with fast forward factor $\alpha=2$ is
shown in Fig.~\ref{fig3}(b).  The external field that is required to
speed up the dynamics by the chosen factor $\alpha=2$ is presented in
Fig.~\ref{fig3}(b3). As expected, $f_{\text{ext}}$ is now a time
dependent field, the amplitude of which decreases monotonically during
the time evolution. In the limit $t\rightarrow\infty$ the external
field converges to that of the original system, since the final
equilibrium state are required to be the same in both cases. In the
Supplemental Material~\cite{SuppMat}, we show a movie of the time
evolution and the required external field in a system which moves
three times faster ($\alpha=3$) than the original system.

As a further example, we conceive a system in which the current 
is the same as in the original system (no speed up, $\alpha=1$),
except for a prescribed additive constant $J_0$. As the divergence of
the constant vanishes, it has no effect on the dynamics of the density
distribution via the continuity equation. Hence the density profiles
of both systems are the same at any time and the current profile in
the new system is $J(t,x)+J_0$, where $J(t,x)$ is the original
current. The required external field that produces such a dynamical
evolution is shown in Fig.~\ref{fig3}(c3), in which we have set
$J_0\sigma\tau=0.5$. The external force field is
again time dependent. The extrema of the force field are shifted with
respect to their original location in the case of the static force
field. This was expected given our above results for the steady state,
Fig.~\ref{fig1}. The amplitude of the force and the magnitude of the
shift vary in a nontrivial way in time, as a result of a delicate
balance between memory effects and the amplitude of the density
modulation. At $t\rightarrow\infty$ the system reaches the same steady state
as that shown in Fig.~\ref{fig1} (orange dashed line).

\vspace{10mm}

\section{Discussion and conclusions}
\label{SECdiscussionConclusions}

We have presented a numerical iterative method to systematically
construct the specific form of the external force field which is
required to drive a prescribed steady state in an overdamped Brownian
many-body system. The same scheme can be used to find the conservative
potential for which a given density profile is in thermodynamic
equilibrium. In equilibrium the method is not restricted to BD
systems.

An alternative
approach has been previously developed for the equilibrium
case~\cite{PRLsuperadiabatic} (also in quantum
systems~\cite{PRLthiele}). Although we have not studied the relative
performance of the two methods systematically against each other,
preliminary tests suggest that the current approach applied to
equilibrium is both faster and more reliable.
Whether the present method can or cannot be extended to quantum
systems is an open and interesting question.

In all cases that we have analysed so far, the iteration processes
have reliably converged.  Nevertheless, if the initial guess for the
external force is very far from the actual force field, it might be
necessary to improve the simple fixed-point iteration scheme presented
here in order to avoid possible divergent trajectories
(i.e.\ sequences of $\fv_{\rm ext}^{(k)}$). Using e.g.\ Anderson
acceleration-like methods should constitute a possible improvement of
the method.  Variations of the presented iterative scheme, such as
e.g.\ defining $\fv_{\rm int}^{(k-1)}=\Fv_{\rm
  int}^{(k-1)}/\rho^{(k)}$ instead of $\fv_{\rm int}^{(k-1)}=\Fv_{\rm
  int}^{(k-1)}/\rho$ in~\eqref{EQiterationStep}, also converge to the
desired external force and might be useful in cases where convergence
issues occur, which might be the case e.g.\ in the vicinity of dynamical
phase transitions.

As the method requires a discretization of the space coordinate,
therefore yields a discretized external force field.  The quality of
the spatial discretization (e.g. size of the bins) is an important
parameter of the method. Although we have shown only one- and
two-dimensional mono-component examples, the extension to
three-dimensional systems and/or mixtures is straightforward.

In all cases, whether time-dependent nonequilibrium,
nonequilibrium steady state, or in equilibrium, the custom flow
method requires to sample the internal force field. Therefore, the
practical implementation for hard-body systems is not as
straightforward as it is in the case of soft interparticle
potentials. For steady-state hard-body systems it might be easier to
extend the 
equilibrium approach of Ref.~\cite{PRLsuperadiabatic} to
nonequilibrium conditions.

The custom flow method allows complete control of the dynamics
of a given system in both steady state and full
nonequilibrium. Possible future applications include the investigation
of time crystals~\cite{TC1,TC2} in BD systems, removal of flow instabilities via the
application of external fields in a controlled manner, and obtaining a
better understanding of memory effects by e.g.\ a systematic analysis
of the external fields required to speed up and/or slow down a given
dynamical process.

\begin{acknowledgments}
This work is supported by the German Research Foundation (DFG) via
SCHM 2632/1-1.
\end{acknowledgments}

\appendix
\section{Sampling the current in Brownian dynamics simulations}
We briefly comment on three different methods to sample the one-body current $\Jv(\rv)$ in Brownian dynamics simulations.
\label{ApCurrent}

{\bf Method 1: Force balance equation}. First, we propose here a new simple method to measure the current
based on the exact one-body force density
balance equation, Eq.~\eqref{EQforceDensityBalance}. This equation provides an expression for $\Jv$ that can be used
to directly sample the current. We need to sample: (i) the internal force density field $\Fv_{\text{int}}$ as an average
over time (in steady state) or over many realizations (in case of time dependent situations), and (ii) the density profile.
Then, using the density profile one can calculate the thermal diffusive term $-k_BT\nabla\rho$.
Finally, the external force density field can either be calculated using the external force and the density
profile ($\Fv_{\text{ext}}=\rho\fv_{\text{ext}}$) or sampled directly during the simulation.

{\bf Method 2: Numerical derivative of the position vector}. The second method, proposed in Refs.~\cite{PRLsuperadiabatic,PRLactive},
is based on calculating the velocity of the $i$th particle, $\vel_i(t)$, via the numerical central derivative of the position vector
\begin{align}
\vel_i(t)=\frac{\rv_i(t+\Delta t)-\rv_i(t-\Delta t)}{2\Delta t}\label{EQnumder}.
\end{align}
Due to the stochastic nature of the motion it is crucial to use the central derivative to properly
compute the velocity of the particles, Eq.~\eqref{EQnumder}. Forward and backward derivatives give
different results that are not consistent with the value of the current obtained by the alternative methods
presented here.

A spatially resolved average of $\vel_i$, Eq.~\eqref{EQnumder}, yields the one-body current profile:
\begin{align}
\Jv(\rv,t)=\left\langle\sum_{i=1}^N\vel_i(t)\delta(\rv_i(t)-\rv)\right\rangle,
\end{align}
where $\langle\cdot\rangle$ indicates an average over either many different realizations
or time in the case of a steady state.

To sample $\vel_i$ more efficiently it is convenient to rewrite Eq.~\eqref{EQnumder} as~\cite{PRLactive}
\begin{align}
\vel_i(t)=\frac{\Delta \rv_i(t-\Delta t)+\Delta\rv_i (t)}{2\Delta t},\label{EQpuufff}
\end{align}
where $\Delta\rv_i (t)=\rv_i(t+\Delta t)-\rv_i(t)$. Plugging Eq.~\eqref{EQBD} into ~\eqref{EQpuufff} results in
\begin{eqnarray}
\vel_i(t) &=& \frac{1}{2\gamma}\left[-\nabla_i u(\rv^N(t-\Delta t))-\nabla_i u(\rv^N(t))+\right.\nonumber\\
&&  \left.{\bf f}_{\rm ext}(\rv_i,t-\Delta t)+{\bf f}_{\rm ext}(\rv_i,t) \right]+\nonumber\\
&& \frac{1}{2\Delta t}[\boldsymbol\eta_i(t-\Delta t)+\boldsymbol\eta_i(t)].
\end{eqnarray}
The spatially resolved average of $\boldsymbol\eta_i(t)$ vanishes at any space point
since $\boldsymbol\eta_i$ is a gaussian random force, and therefore this average correlates the random force at time $t$ with
the position at the same time $t$. In contrast, it is important to realize that the spatially resolved average of $\boldsymbol\eta_i(t-\Delta t)$
does not vanish in general, since it correlates the random force at time $t-\Delta t$ with the current position at time $t$.

{\bf Method 3: Continuity equation}. The continuity equation, Eq.~\eqref{EQcontinuity} provides an alternative
route to compute the current in non-steady state situations, as shown in Ref.~\cite{PRLsuperadiabatic}.
Having sampled the density profile at different times $t$ and $t'$, we can compute
the numerical time derivative of the density profile, which must be equal to the divergence of the current. 
In effectively one-dimensional systems the result can be integrated in space and yields the one-body current profile (line
integrals or other inversion methods are required in a higher dimensional space).
This method yields the current profile up to an additive constant. If the actual value of the current at a given
space point is known, then one can easily determine the missing additive constant. For instance, if the system is in contact
with a hard wall the current at the hard wall must vanish. To use this method  we need to sample $\rho$ at two times $t$ and $t'$
separated by a time interval $\Delta_t$. In our experience a value $\Delta_t\approx 10^2\Delta t$ with $\Delta t$ the time step
of the BD simulation provides good results.

We have checked the three methods presented above give the same one-body current profile within the inherent numerical accuracy 
of each procedure.

\section{Numerical details}
We provide details of our precise implementation of the iterative 
scheme.
\label{ApB}
\begin{figure}
\includegraphics[width=0.80\columnwidth]{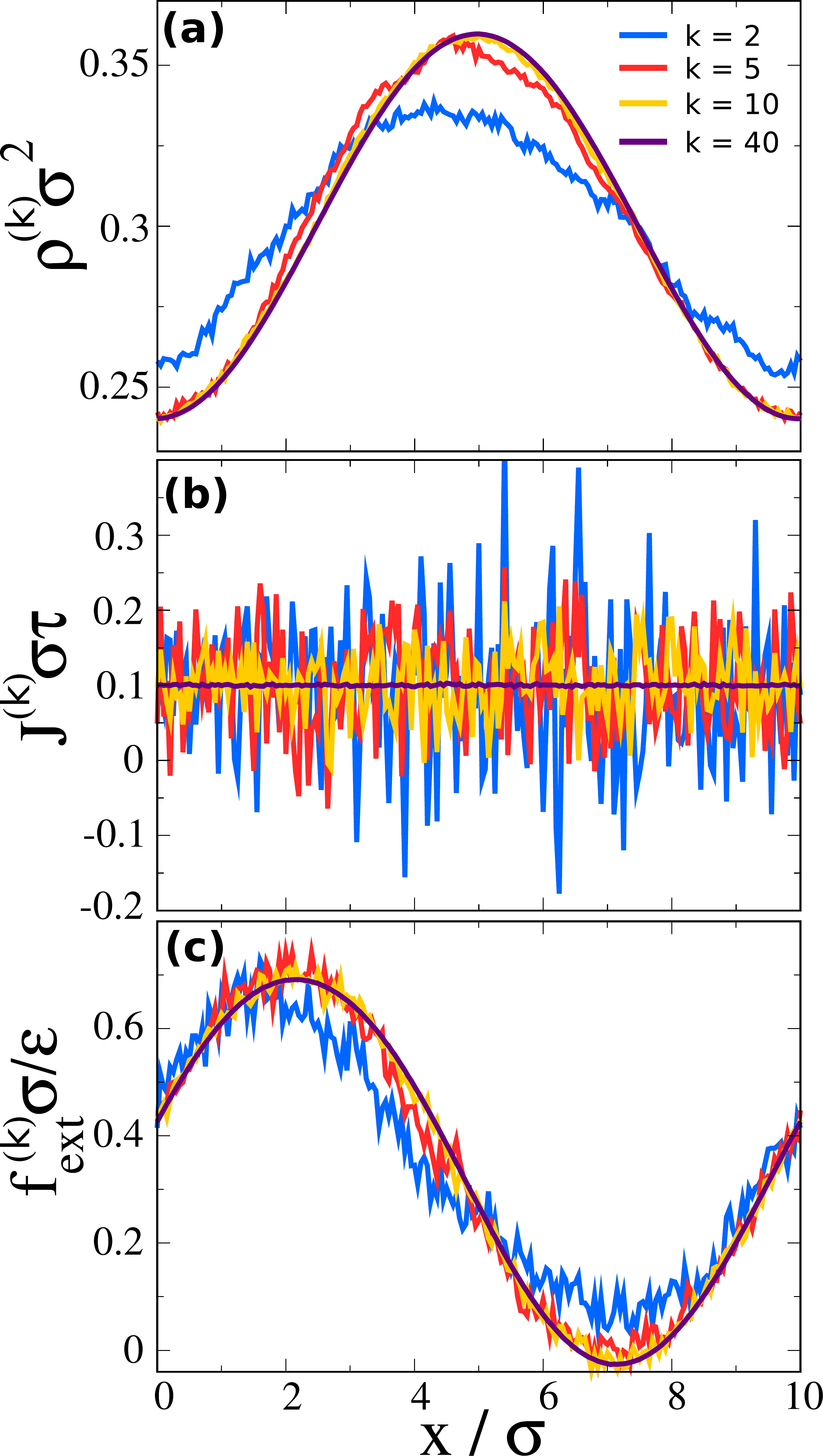}
\caption{One-body density profile (a), one-body current (b), and
  external force field (c) as a function of the $x$-coordinate for different
  number of iterations, $k$, as indicate in the legend of (a). The
  target current is set to $J_0\sigma\tau=0.1$.  The target density
  profile is $\rho_\tar(x)=c_1\sin^2(\pi x/L)+c_2$ with
  $c_1\sigma^{2}=0.12$ and $c_2\sigma^2=0.24$, which is practically
  identical to the sampled density profile after $40$ iterations
  (violet solid line). Two-dimensional system with $N=30$
  particles in a periodic box of side $L/\sigma=10$ at temperature
  $k_BT/\epsilon=1$. The data has been obtained by averaging over $25$
  BD realizations. The total simulation time of iteration $k$ of one
  BD realization is set to $\tau_k/\tau=\tau_02^{(k-1)/3}$, with
  $\tau_0/\tau=100$. The bin size is $\Delta x/\sigma = 0.05$.}
\label{figA1}
\end{figure}
Each iteration step \eqref{EQiterationStep} of the nonequilibrium
inversion method requires carrying out one BD simulation run for the
given parameters and given external force field. Before acquiring the
data, we let the system reach a steady state during $10^2\tau$. Then,
at each iteration $k$, we sample the internal force density during a
given sampling time $\tau_k$. The sampling time has a direct impact on
both the statistical quality of the sampled internal force field
(which is required for the next iteration) as well as on the
performance of the method.  Instead of using the same sampling time at
each iteration, we find it preferable to start with short simulation
runs and increase the run length at every iteration.  Hence, at
iteration $k$, we fix the sampling simulation time $\tau_k$ to
\begin{align}
  \tau_k=\tau_02^{(k-1)/3},
  \label{eqtime}
\end{align}
i.e., we double the runlength every three iterations. The total time
of the first iteration is set to $\tau_0/\tau=100$. Finally, we
average over several ($10-100$) realizations of the iteration scheme,
Eq.~\eqref{EQiterationStep}, to improve the statistics.

\begin{figure}
\includegraphics[width=0.80\columnwidth]{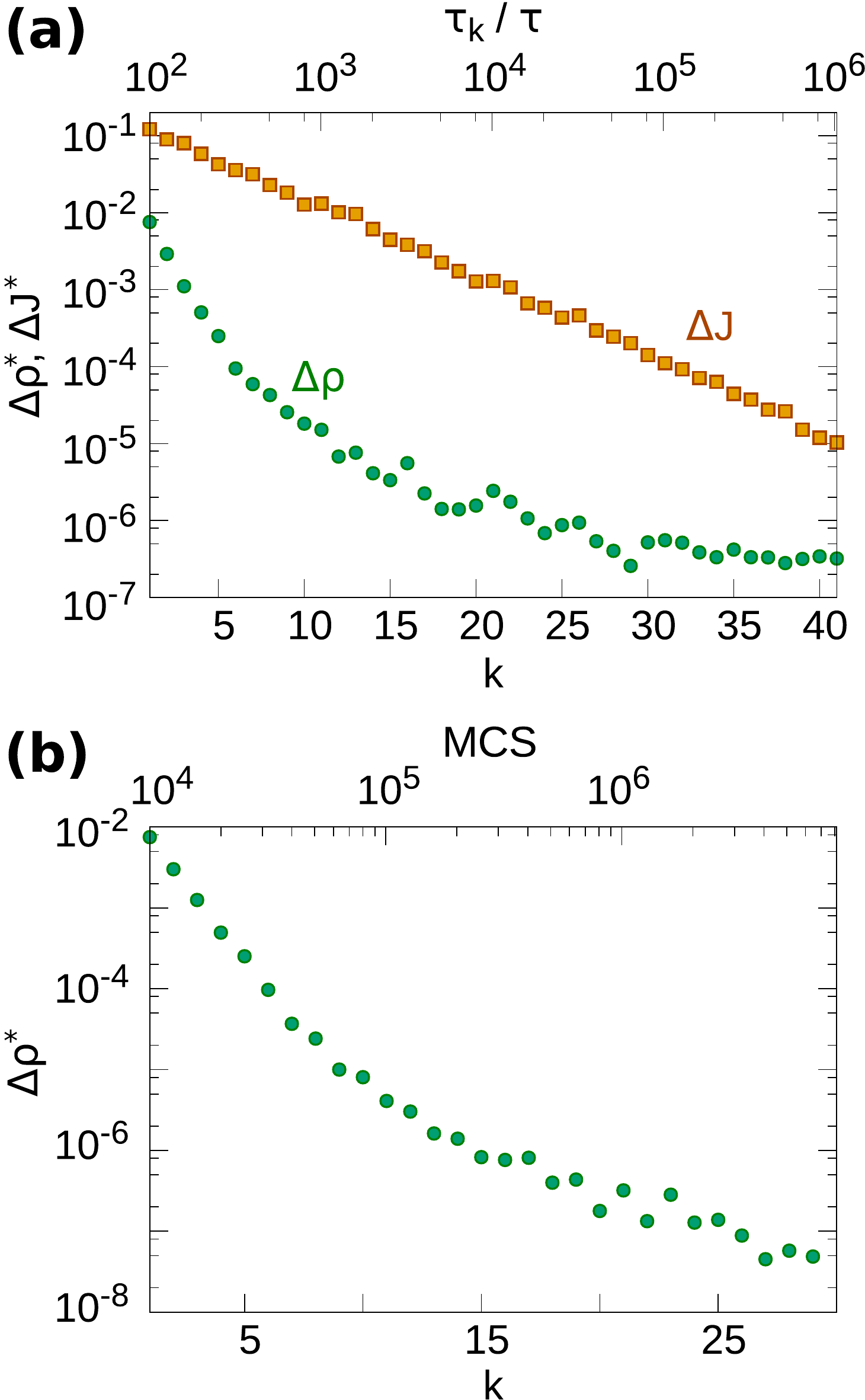}
\caption{(a) Scaled error of the density profile
  $\Delta\rho^*=\Delta\rho L^2\sigma^2$ (green circles) and of the current
  profile $\Delta J^*=\Delta JL^2\tau^2$ (yellow squares) as a function
  of the iteration number $k$ (bottom axis) and the scaled simulation
  time of the iteration $\tau_k/\tau$ (upper axis) in non-equilibrium steady state using BD simulations.
  (b) Scaled error of the density profile $\Delta\rho^*=\Delta\rho L^2\sigma^2$ 
  as a function of the iteration number $k$ (bottom axis) and the number of Monte Carlo steps 
  at iteration $\tau_k/\tau$ (upper axis) in equilibrium using MC simulations. 
  In both panels, the data has been
  obtained by averaging over $25$ realizations. Note the logarithmic
  scale of the vertical axis and of the upper horizontal axis.
  }
\label{figA2}
\end{figure}

Fig.~\ref{figA1} illustrates the iterative process for the effectively
one-dimensional system with target profile given by
Eq.~\eqref{targetrho1} and target current $J_0\tau\sigma=0.1$.  Less
than 10 iterations suffice to get a very good estimate of the external
force, and after $k=40$ iterations the sampled density, and current
profiles are almost indistinguishable from the corresponding target
profiles.  The results have been obtained by averaging over $25$ BD
realizations of the iterative scheme. We show in Fig.~\ref{figA2}(a)
the evolution of the error of the density and the current profile
during the iterative process, cf.\ Eqs.~\eqref{errorDensity} and
\eqref{errorCurrent}.

For the equilibrium situation of the effectively one-dimensional
system shown in Fig.~\ref{fig1} ($J_0=0$) we have used Monte Carlo
simulations to implement the iterative scheme, cf.~\eqref{iteeq}
and~\eqref{ulala}. For completeness, we also show the efficiency of
the method in equilibrium in Fig.~\ref{figA2}(b), where we plot the
difference between the sampled and the error in the density profile as
a function of the number of iterations and the number of Monte Carlo
sweeps (MCS).  Each MCS is an attempt to sequentially and individually
move all the particles in the system.  We find it convenient to
increase the number of MCS during the iterative process. We begin with
$10^4$ MCS at iteration $k=1$, and increase the number of MCS after
every iteration such that it doubles every three iterations. Before
acquiring data we equilibrate the system by running $10^4$ MCS.  As in
the non-equilibrium steady state case, we improve the statistics by
averaging over $25$ realizations.

\end{document}